\documentclass[conference]{IEEEtran}
\IEEEoverridecommandlockouts
\usepackage{cite}
\usepackage{amsmath,amssymb,amsfonts}
\usepackage{algorithmic}
\usepackage{graphicx}
\usepackage{textcomp}
\usepackage{xcolor}
\usepackage{url}
\usepackage{xurl}
\def\BibTeX{{\rm B\kern-.05em{\sc i\kern-.025em
 b}\kern-.08em
    T\kern-.1667em\lower.7ex\hbox{E}\kern-.125emX}}

\ifCLASSOPTIONcompsoc
  \usepackage[nocompress]{cite}
\else
  \usepackage{cite}
\fi
\ifCLASSINFOpdf
\else
\fi

\usepackage{graphicx}

\DeclareRobustCommand*{\IEEEauthorrefmark}[1]{%
  \raisebox{0pt}[0pt][0pt]{\textsuperscript{\footnotesize #1}}%
}

\begin{document}

\title{Advanced Printed Sensors for Environmental Applications: A Path Towards Sustainable Monitoring Solutions \\

}

\author{\IEEEauthorblockN{Nikolaos Papanikolaou\IEEEauthorrefmark{1, 2},
Doha Touhafi\IEEEauthorrefmark{1},
Jurgen Vandendriessche\IEEEauthorrefmark{1}, 
Danial Karimi\IEEEauthorrefmark{2},
Sohail Fatimi\IEEEauthorrefmark{2}, \\
Gianluca Cornetta\IEEEauthorrefmark{1,3},and
Abdellah Touhafi\IEEEauthorrefmark{1,2},~\IEEEmembership{Fellow,~IEEE}}
\IEEEauthorblockA{\IEEEauthorrefmark{1}Lumency BVBA– Brussels, Belgium}
\IEEEauthorblockA{\IEEEauthorrefmark{2}Vrije Universiteit Brussel– Brussels,Belgium}
\IEEEauthorblockA{\IEEEauthorrefmark{3} San Pablo-CEU University– Madrid, Spain}
}



\maketitle

\begin{abstract}
Printed sensors represent a transformative advancement in sensor technology, utilizing innovative printing techniques to create flexible, cost-effective, and highly customizable sensing devices. Their versatility allows integration into numerous applications across diverse fields such as monitoring a wide range of environmental factors e.g. air and water quality, soil conditions, and atmospheric changes among others. These sensors demonstrate high sensitivity and accuracy in detecting pollutants, temperature variations, humidity levels, and other critical parameters essential for environmental assessment and protection. 

The adaptability of printed sensors to operate under various environmental conditions expands their applicability, making real-time monitoring feasible in both urban and remote settings. This is particularly valuable in areas where traditional monitoring systems are impractical or too costly to implement, thus broadening the scope of environmental monitoring. Their lightweight and flexible design further enhances their suitability for deployment in challenging environments, including remote and hard-to-reach locations. The integration of printed sensors with wireless communication technologies enables the formation of comprehensive monitoring networks, facilitating continuous data collection and real-time data transmission. These networks aggregate information from multiple sensors, improving situational awareness and supporting data-driven decision-making in areas like environmental management, policy development, and disaster response. 

Research efforts continue to focus on enhancing the performance of printed sensors, improving their sensitivity, and ensuring long-term reliability. Advances in material science and printing techniques are driving improvements in sensor capabilities, leading to greater accuracy and durability. This paper provides a comprehensive review of recent developments and challenges in printed sensor technology, discussing their applications in environmental monitoring and highlighting their potential to address pressing environmental issues. By exploring advancements in sensor materials, fabrication methods, and integration with digital technologies, this paper aims to underscore the transformative potential of printed sensors in revolutionizing environmental monitoring practices. Ultimately, these insights contribute to the sustainable management and protection of natural resources, aligning with global efforts to promote a more sustainable future.

\end{abstract}


\section{Introduction}
Considering escalating environmental challenges, the demand for innovative and sustainable monitoring solutions is more urgent than ever. Sensors are now widely employed to detect and track environmental changes, with their applications expanding rapidly across various fields [1][2]. These sensors are often installed on or connected to different objects and can be operated remotely via control engineering, offering flexibility in monitoring diverse environments [3]. 

Until the end of the last decade, single-crystal silicon was the dominant semiconductor material for developing sensor substrates, as shown in Fig.1. Silicon-based sensors offer distinct advantages, such as cost-effectiveness in terms of large-scale production, high sensitivity, and low leakage currents, thanks to the material’s high potential barrier. Despite their widespread use in large-scale sensor production, silicon sensors come with notable drawbacks, including high initial fabrication costs, significant power consumption, susceptibility to mechanical damage, and inherent stiffness. These limitations have led to an increasing preference for flexible  prototypes for sensing in various applications [4][5]. This review paper explores the potential of advanced printed sensors as a transformative approach to environmental monitoring, providing a more sustainable and efficient alternative to conventional technologies. 

\begin{figure}[htbp]
\centerline{\includegraphics[scale=0.55]{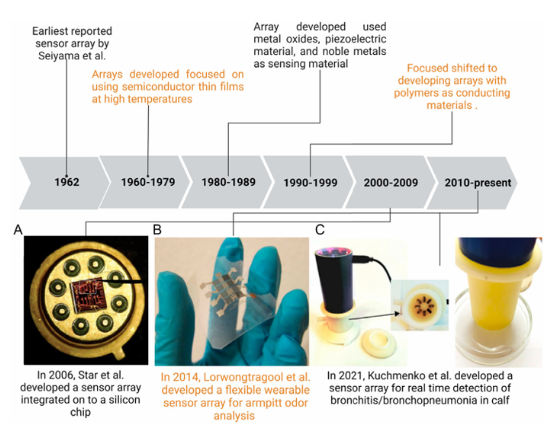}}
\caption{Evolution of sensor array timeline (1962-present)[6].}
\label{evolution}
\end{figure}

Printed sensors leverage advancements in material science and printing technologies to offer a versatile and cost-effective solution [7][8]. Unlike traditional sensors, these can be fabricated on flexible substrates, making them suitable for a wide range of applications, including air and water quality monitoring, soil health assessments, and many more applications. The ability to mass-produce printed sensors through environmentally friendly processes further enhances their appeal in promoting sustainable practices [9]. 

This paper is structured into several sections that explore various aspects of printed sensors. Section 2 is divided into two parts. The first part focuses on the materials used in the development of printed sensors, discussing recent innovations and their impact on both sensor performance and sustainability. In addition, the second part examines advancements in printing techniques and sensor fabrication, highlighting the progress achieved as well as the challenges encountered in this rapidly evolving field. Moving to section 3, the paper delves into the specific applications of printed sensors in environmental monitoring, showcasing their versatility and effectiveness. The comparison between existing technologies while outlining the prospects for printed sensors, offering insights into their potential to revolutionize environmental monitoring practices is better analysed in section 4 . In section 5, the challenges and limitations of printed sensors are addressed, providing a balanced perspective on their current capabilities and areas for improvement. Finally, the conclusion of the paper is presented in section 6 by summarizing key findings and emphasizing the importance of adopting advanced printed sensors as essential tools for achieving sustainable environmental monitoring solutions.

\section{Technological Developments and Challenges}

Printed sensors utilize a range of materials and fabrication techniques, each offering unique benefits and drawbacks.  

Polymers like PolyEthylene Terephthalate (PET) and PolyImide (PI) are valued for their flexibility, enabling them to adapt to complex surfaces and integrate seamlessly into diverse devices. Their printability simplifies manufacturing, while their mechanical strength supports delicate components. However, polymers are sensitive to high temperatures and may degrade over time when exposed to harsh environments ( with high temperature and humidity conditions), raising concerns about their long-term reliability [10][11].  

Metals such as copper and aluminum have excellent electrical conductivity, essential for high signal fidelity. Their durability enhances reliability in applications requiring structural integrity. However, metals are limited by their rigidity and weight, making them less suitable for applications that prioritize flexibility and lightweight designs [4][9][12].  

\begin{figure}[htbp]
\centerline{\includegraphics[scale=0.85]{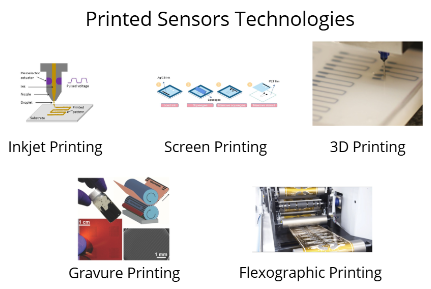}}
\caption{Printed Sensors Technologies.}
\label{printsenstech}
\end{figure}

Ceramics offer high-temperature stability and chemical resistance, ideal for harsh industrial environments. Yet, ceramics are brittle and prone to fracture under stress, and their complex manufacturing processes increase costs, limiting their scalability [11]. 

Nanomaterials, like graphene and carbon nanotubes, provide exceptional sensitivity and enable sensor miniaturization, making them ideal for precision applications. However, challenges in scaling up production and ensuring long-term stability have hindered their widespread adoption [10].  

The development of printed sensors relies on several fabrication techniques, each with distinct characteristics, as presented in Fig. 2. Inkjet printing is known for its cost-effectiveness and high precision, but it is limited to specific inks and is generally slower than other methods, affecting production efficiency [13]. Screen Printing is suitable for large-scale production, allowing thick material deposition. However, it struggles with lower resolution and complex processes that require multiple steps [14]. 3D Printing (additive manufacturing) excels in customization, enabling direct integration of sensors into printed objects, but its slower speed and higher initial costs can be a barrier [10]. Flexographic Printing offers high speed and flexibility for large production runs, but often yields lower-quality prints and involves a time-consuming setup [15]. Finally, gravure printing provides high-quality, consistent results and durability for large-volume applications, yet suffers from high initial costs and limited substrate compatibility [16]. 
 
Each technique faces unique challenges: Inkjet Printing contends with ink formulation and ensuring sensor durability in various conditions. Screen Printing struggles with material compatibility and achieving precision. 3D Printing requires developing printable, functional materials and managing post-processing. Flexographic Printing must optimize ink adhesion and sensor robustness, while Gravure Printing deals with complex setups and material restrictions [13-16].  

Common challenges across all methods include achieving comparable sensitivity and accuracy to traditional sensors, ensuring environmental stability, developing new printable materials, and integrating sensors into existing systems. Overcoming these hurdles is crucial for advancing printed sensor technology and unlocking its full potential in diverse applications [17][18]. 

\section{Applications of Printed Sensors for Environmental Monitoring}

Printed sensors are emerging as a transformative technology for environmental monitoring, primarily due to their cost-effectiveness, adaptability, and suitability for large-scale production, as shown in Fig. 3. One of the primary applications of printed sensors is in air quality monitoring, where they can detect a wide range of harmful pollutants, such as carbon monoxide (CO), nitrogen dioxide (NO2), and particulate matter (PM). Precise monitoring of these pollutants is critical for managing air quality, particularly in densely populated urban areas. Moreover, integrating printed sensors into display systems provides real-time data, facilitating immediate decision-making and enabling authorities to issue timely warnings when pollution levels exceed regulatory thresholds [19].   

\begin{figure}[htbp]
\centerline{\includegraphics[scale=0.5]{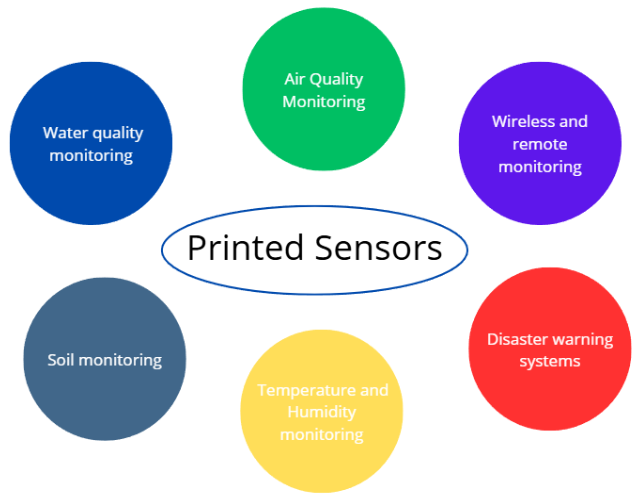}}
\caption{Environmental applications of screen printed sensors.}
\label{printedsensorsapp}
\end{figure}

Printed sensors are also essential for monitoring water quality, as illustrated in Fig. 4. Screen-printed electrochemical sensors, for example, are highly effective in detecting heavy metal ions like lead and mercury, which pose serious risks to both human health and the environment [20]. Additionally, printed nitrate sensors are valuable tools for monitoring water safety and promoting sustainable agricultural practices by minimizing the overuse of fertilizers, which can cause nutrient pollution in water bodies [21].

\begin{figure}[htbp]
\centerline{\includegraphics[scale=0.35]{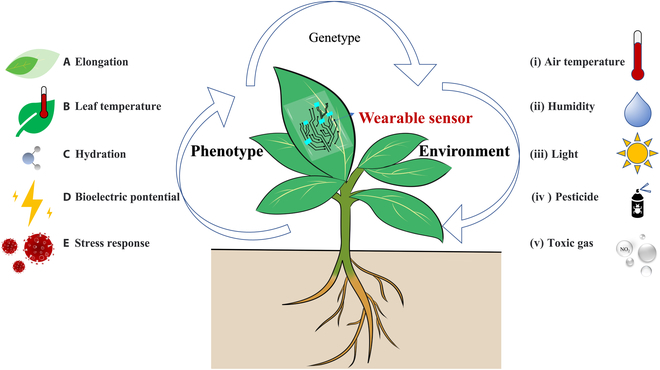}}
\caption{Printed sensor in agriculture applications [22].}
\label{printedagri}
\end{figure}

In agricultural settings, printed sensors are also used for soil monitoring, as depicted in Fig. 5. Soil moisture sensors provide accurate measurements of moisture levels, which are essential for optimizing irrigation management and conserving water [21]. Furthermore, printed sensors that detect soil nutrient concentrations enable farmers to apply fertilizers more efficiently, thereby enhancing crop yields while minimizing the environmental impact of excessive fertilizer use.

\begin{figure}[htbp]
\centerline{\includegraphics[scale=0.5]{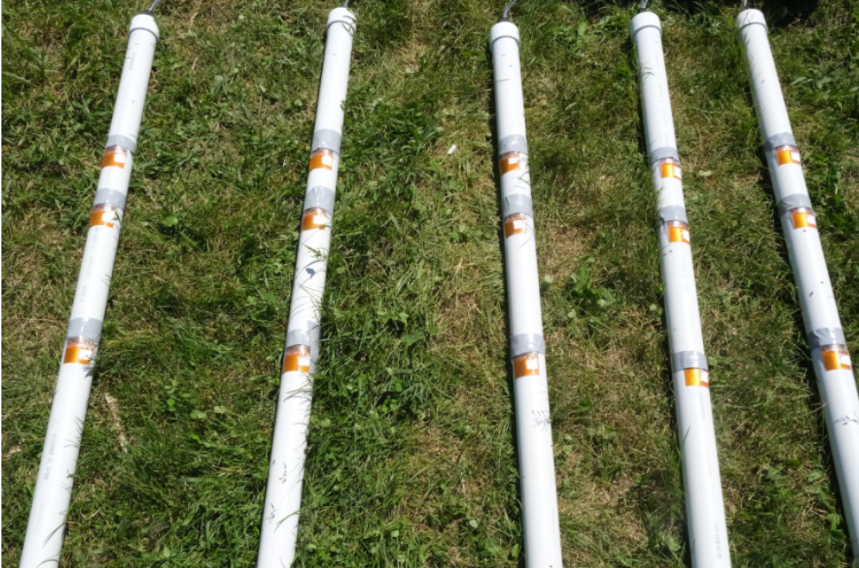}}
\caption{Printed Sensors in Soil Applications [23].}
\label{soilapp}
\end{figure}

Printed sensors are also widely used for temperature and humidity monitoring, as shown in Fig. 6, aiding in climate control and energy management across various settings. From greenhouse agriculture to building automation, these sensors offer flexibility and ease of integration, making them ideal for maintaining optimal environmental conditions [7].

\begin{figure}[htbp]
\centerline{\includegraphics[scale=0.4]{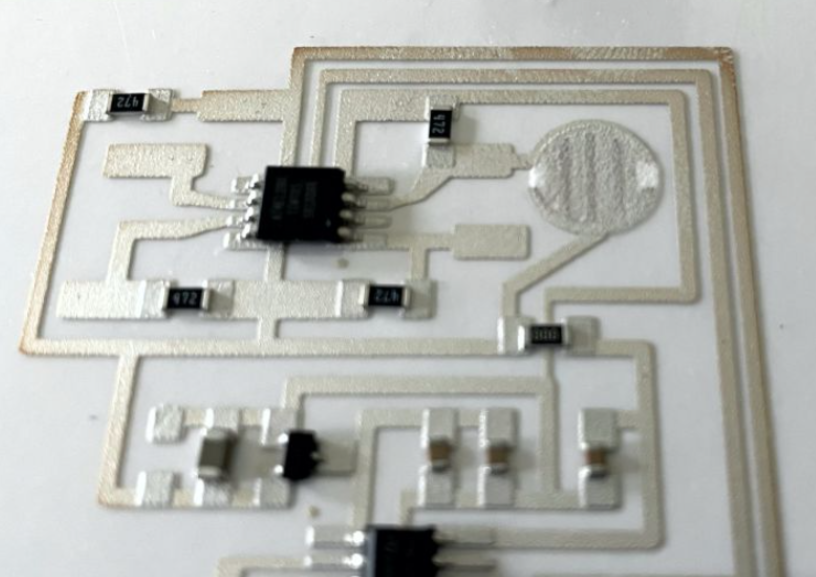}}
\caption{Printed Humidity Sensor [24].}
\label{humsens}
\end{figure}

Beyond these applications, printed sensors contribute significantly to disaster warning systems, as depicted in Fig. 7. By detecting early signs of natural disasters such as earthquakes and tsunamis, these sensors provide critical data that enable timely evacuations and help minimize potential loss of life and property damage. Their ability to rapidly identify environmental changes makes them valuable assets for enhancing community resilience against natural hazards [7]. 

\begin{figure}[htbp]
\centerline{\includegraphics[scale=0.5]{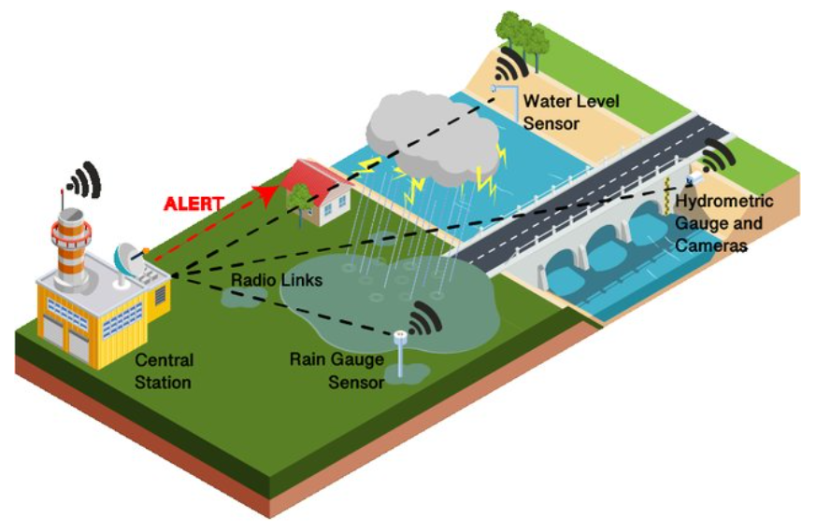}}
\caption{Sensors in disaster warning systems [25].}
\label{disaswarnsens}
\end{figure}

The capability of printed sensors to support wireless and remote monitoring further extends their utility to remote and hard-to-reach areas [19]. This feature allows for the continuous collection of environmental data over large geographical regions, enabling a comprehensive understanding of environmental trends and a proactive response to emerging threats. Additionally, printed sensors find use in biosensors for ecosystem health, greenhouse gas monitoring (such as methane and $CO_{2}$ detection), noise pollution monitoring, food quality control, smart agriculture, energy and waste management, ocean monitoring, forest fire detection, and numerous other applications [7].  

Printed sensors’ potential for integration into a broad range of applications underscores their role as critical tools in advancing sustainability and public safety. Continued research and development in this field will be essential to overcome existing challenges and unlocking the full potential of printed sensors for environmental monitoring. 
\section{Comparison of Existing Technologies and Future Prospects}

From a technical perspective, printed sensors offer significant advantages. One of their greatest benefits is their cost-effective production, as they are considerably cheaper to manufacture compared to traditional silicon-based sensors. This reduction in production costs makes printed sensors more accessible for a wide range of applications, as illustrated in Fig. 8 [26]. Additionally, their versatility is a key strength, as they can be printed on various substrates, including flexible and stretchable materials, broadening their usability across diverse fields [27]. Another technical advantage is their support for rapid prototyping. The digital nature of the printing process allows for quick design modifications, significantly shortening development time [17][28]. Furthermore, printed sensors are highly scalable, enabling mass production while maintaining consistent quality, a crucial factor for large-scale deployments [7]. 

\begin{figure}[htbp]
\centerline{\includegraphics[scale=0.7]{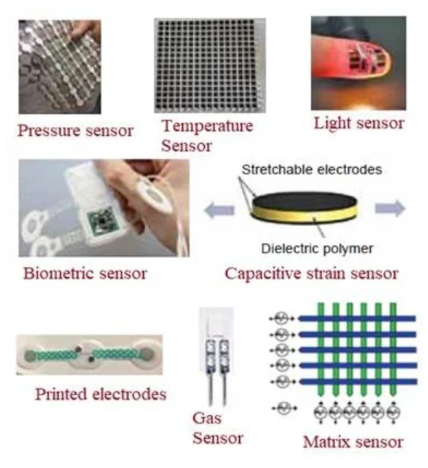}}
\caption{Different types of printed sensors[27].}
\label{typesprint}
\end{figure}

From an environmental perspective, printed sensors present several advantages. The printing process generates less waste compared to conventional manufacturing techniques, reducing the overall environmental footprint [19]. This is further enhanced by the difference in energy consumption of the production processes. Printed sensor have a more energy efficient production process which also improves their environmental impact. Moreover, recyclability is another important benefit, as many printed sensors can be designed with end-of-life recycling in mind, enhancing their sustainability. 

Looking ahead, printed sensors are increasingly being integrated with Internet of Things (IoT) devices, enabling real-time data collection and analysis. This trend is expected to accelerate as the demand for connected devices grows [29]. Additionally, ongoing advancements in material science are poised to improve the performance and durability of printed sensors, addressing some of their current limitations. Miniaturization is also emerging as a key trend, driven by the rising demand for wearable and portable devices that require smaller, more compact sensors. Lastly, there is a growing emphasis on sustainability, with researchers focusing on developing eco-friendly materials and processes to enhance the environmental benefits of printed sensors [30][31]. 

In the field of environmental monitoring, printed sensors hold significant promise. Their cost-effectiveness and flexibility make them ideal for monitoring air and water quality, detecting pollutants, and tracking environmental changes in real-time. The ability to produce these sensors at scale and at a lower cost suggests they could be widely deployed, offering a practical solution for comprehensive environmental monitoring [7].

\section{Challenges and Limitations}

Printed sensors have shown great promise as a transformative technology in various fields due to their cost-effectiveness, flexibility, and ease of production. However, their widespread adoption is currently hindered by several limitations and challenges, particularly in terms of reliability, accuracy, and durability.  
  
One of the primary concerns surrounding printed sensors is their reliability. Traditional sensors, typically composed of silicon or other robust materials, have long been valued for their consistent performance over time. In contrast, printed sensors, which are often fabricated using organic or composite materials, may suffer from variability in performance. The reliability of these sensors is influenced by factors such as the formulation of the inks, the quality of the printing process, and the substrates used. Such variations can lead to inconsistent performance, particularly in large-scale production runs. As a result, achieving consistent reliability across a wide range of applications remains a significant challenge in the development of printed sensors [7][11].

Accuracy is another critical issue that printed sensors face, especially when compared to conventional sensors. The precision of these sensors can be influenced by various factors, including the resolution of the printing process, the uniformity of the materials used, and the stability of the sensor’s response over time. In applications that require high accuracy, such as medical applications or precise environmental monitoring, the current generation of printed sensors may not meet the stringent standards required. Enhancing the accuracy of printed sensors is therefore a focal point of ongoing research, with efforts aimed at improving both the materials and the printing techniques to deliver more reliable and precise measurements [32][33].

Durability presents yet another challenge, particularly when printed sensors are deployed in harsh or adverse environmental conditions. Traditional sensors are often designed to withstand extreme temperatures, high humidity, and mechanical stress. In contrast, printed sensors are more susceptible to degradation under such conditions. For example, exposure to UltraViolet (UV) light, moisture, or certain chemicals can cause the materials in printed sensors to deteriorate, leading to a loss of functionality over time. To enhance the durability of these sensors, researchers are exploring new materials and protective coatings that can help them withstand extreme environmental conditions, thus improving their longevity and performance [11][34].

In addition to these technical challenges, the range of materials that can currently be used in printed sensors is somewhat limited. Although significant progress has been made in developing printable conductive inks and substrates, there remains a need for materials that offer higher performance while maintaining environmental stability. Ongoing research is focused on developing new composites and nanomaterials that could enhance both the functionality and durability of printed sensors, expanding their potential applications [7].

Another critical issue is the integration of printed sensors into existing systems. These sensors must be compatible with other electronic components, which can be challenging given the diversity of materials and processes used in their fabrication. Additionally, there is currently a lack of standardized testing and quality assurance protocols across the industry, which can lead to inconsistencies in sensor performance. Establishing industry-wide standards for the production, testing, and integration of printed sensors will be essential for ensuring consistent quality and performance across different applications [32][35].

\begin{figure}[htbp]
\centerline{\includegraphics[scale=0.3]{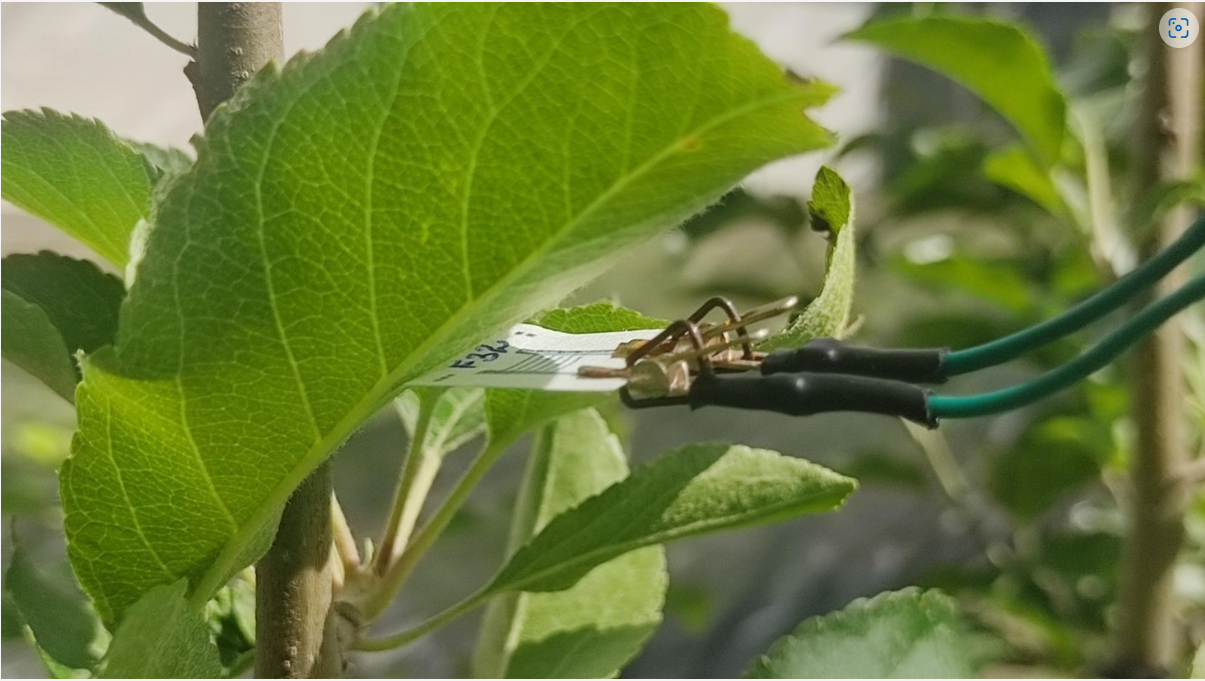}}
\caption{Printed Sensors in environmental application [36].}
\label{plantsens}
\end{figure}

Addressing these challenges requires a multidisciplinary approach that combines advances in material science, engineering, and manufacturing processes. Several promising directions for future development include creating new materials to enhance the performance and stability of printed sensors, innovating printing technologies for higher resolution and better material control, and conducting rigorous environmental testing to identify and mitigate potential failure modes such as material degradation, mechanical failure, electrical malfunctions, adhesion problems, and calibration drift. Additionally, efforts to standardize production and testing protocols will be crucial for ensuring that printed sensors can consistently meet the demands of various applications [10][11]. 

\section{Conclusion}
The development and implementation of advanced printed sensors for environmental applications marks a pivotal advancement towards sustainable monitoring solutions. These sensors offer numerous advantages, including cost-effectiveness, flexibility, and versatility in deployment across various environments. By leveraging innovative materials and printing techniques, these sensors can deliver real-time, accurate data crucial for effective environmental monitoring and management. Their integration into existing frameworks enhances our ability to promptly detect and respond to environmental changes. This advancement not only supports environmental protection efforts but also promotes the sustainable use of resources. Future research should prioritize improving the durability and sensitivity of these sensors, as well as exploring new applications and integration methods to fully realize their potential. Overall, advanced printed sensors hold great promise for contributing to a more sustainable and environmentally conscious future. 

\section*{Acknowledgment}

This paper was written with the valuable assistance of colleagues from Lumency and the VUB University. I would like to personally thank Professor Abdellah Touhafi for his support and help.


\end{document}